\documentclass[letterpaper,reprint,aps,pra,superscriptaddress,showpacs,amsmath,amssymb,floatfix]{revtex4-1}

\usepackage{graphicx}
\usepackage{txfonts}
\usepackage{bm}

\usepackage{braket}
\usepackage{color}
\allowdisplaybreaks[1]

\graphicspath{{./Figure/}}

\begin{document}

\title{Purification of photon subtraction from continuous squeezed light by filtering}
\author{Jun-ichi Yoshikawa}
\email{yoshikawa@ap.t.u-tokyo.ac.jp}
\affiliation{Department of Applied Physics, School of Engineering, The University of Tokyo, 7-3-1 Hongo, Bunkyo-ku, Tokyo 113-8656, Japan}
\affiliation{Quantum-Phase Electronics Center, School of Engineering, The University of Tokyo, 7-3-1 Hongo, Bunkyo-ku, Tokyo 113-8656, Japan}
\author{Warit Asavanant}
\affiliation{Department of Applied Physics, School of Engineering, The University of Tokyo, 7-3-1 Hongo, Bunkyo-ku, Tokyo 113-8656, Japan}
\author{Akira Furusawa}
\email{akiraf@ap.t.u-tokyo.ac.jp}
\affiliation{Department of Applied Physics, School of Engineering, The University of Tokyo, 7-3-1 Hongo, Bunkyo-ku, Tokyo 113-8656, Japan}

\begin{abstract}
Photon subtraction from squeezed states is a powerful scheme to create good approximation of so-called Schr\"odinger cat states. 
However, conventional continuous-wave-based methods actually involve some impurity in squeezing of localized wavepackets, even in the ideal case of no optical losses. 
Here we theoretically discuss this impurity, by introducing \textit{mode-match} of squeezing. 
Furthermore, here we propose a method to remove this impurity by filtering the photon-subtraction field. 
Our method in principle enables creation of pure photon-subtracted squeezed states, which was not possible with conventional methods. 
\end{abstract}

\date{\today}

\maketitle

\section{Introduction}

Coherent states $\ket{\alpha}$ are the quantum states most close to the classical waves with complex amplitude $\alpha$, and cheap resources available from laser light. 
However, when they are superposed as $\ket{\alpha}+e^{i\theta}\ket{-\alpha}$, they become highly nonclassical, non-Gaussian states, often referred to as Schr\"odinger cat states. 
Note that the normalization factor is ignored here and in the following when not necessary. 
Coherent-state superpositions (CSSs) $c_+\ket{\alpha}+c_-\ket{-\alpha}$ are one of promising implementations of qubits, enabling quantum computation \cite{Ralph.PRA2003}. 
In addition, measurements in some CSS bases can provide better discrimination of coherent states \cite{Ban.PRA1997}, which can boost capacities of classical communication.

Although it is currently hard to prepare general CSS qubits with large amplitude as traveling light wavepackets, it is known that photon-subtracted squeezed states well approximate plus or minus cat states $\ket{\alpha}\pm\ket{-\alpha}$ when the amplitude $\lvert\alpha\rvert$ is not large (typically $\lvert\alpha\rvert \leq 1.2$) \cite{Dakna.PRA1997,Lund.PRA2004}. 
Based on this theory, photon-subtraction experiments are conducted: 
Initially, one-photon subtraction is succeeded by using a pulsed laser \cite{Ourjoumtsev.Science2006}, and then also by using a continuous wave (CW) laser \cite{Neergaard.PRL2006,Wakui.OptExpr2007}. 
Later, two-photon subtraction \cite{Takahashi.PRL2008} and three-photon subtraction \cite{Gerrits.PRA2010} are also successfully demonstrated. 
Starting from a squeezed vacuum state which is superposition of even-number states, subtraction of an odd number of photons results in superposition of odd-number states approximating a minus cat state, while subtraction of an even number of photons results in superposition of even-number states approximating a plus cat state. 
Furthermore, the photon-subtraction scheme is extended to generation of parity qubits \cite{Neergaard.PRL2010}. 
The amplitude of cat states can be enlarged with conditional methods \cite{Lund.PRA2004,Sychev.NPhoton2017}. 
Cat states are resources for teleamplification of coherent states \cite{Neergaard.NPhoton2013}. 
Hybridization of coherent-state qubits and number-state qubits is also demonstrated \cite{Jeong.NPhoton2014, Morin.NPhoton2014}.

In particular, approximative minus cat states obtained by one-photon subtraction are actually squeezed single-photon states, having a negative region around the origin of the Wigner function. 
Negative regions in the Wigner function are a clear evidence of strong nonclassicality of the quantum states. 
However, in real experiments, optical losses contaminate the odd parity ($P_{2k}=0$ for all $k\in\mathbb{N}$) of the minus cat states, degrading the negative value at the origin $W(0,0)=(1/\pi)\sum_{k=0}^{\infty}(P_{2k}-P_{2k+1})$ from the ideal $-1/\pi$. 
Here, $P_n$ is the $n$-photon component of a single-mode quantum state in a given wavepacket mode. 
The best negativities (without correction of any losses) of about $-0.17$ of minus cat states are demonstrated with the CW scheme and utilized as input states of a quantum teleporter \cite{Lee.Science2011} or a squeezing gate \cite{Miwa.PRL2014}. 
Based on these successes, we concentrate on the CW scheme in this paper. 
An advantage of the CW scheme is high interference visibilities of cat states with local oscillators of homodyne detection for quantum tomographic characterization.

Photon subtraction is a conditional, nonunitary operation, achieved by tapping a small portion of the initial state with an asymmetric beamsplitter and measuring it with a photon detector, as explained in Sec.~\ref{sec:Basics}. 
When a photon is detected, the photon subtraction is succeeded, and the heralded photon-subtracted state exists in some wavepacket, localized in the time domain around the heralding signal \cite{Molmer.PRA2006}. 
However, here we pay attention to the fact that the initial squeezed vacuum state in such a wavepacket is generally in a mixed state, owing to the nonflat spectrum of squeezing produced by an optical parametric oscillator (OPO). 
This impurity of the initial squeezed states would remain as impurity of the heralded cat states in some form \cite{Takeoka.PRA2008}.

\begin{figure}[tb]
\centering
\includegraphics{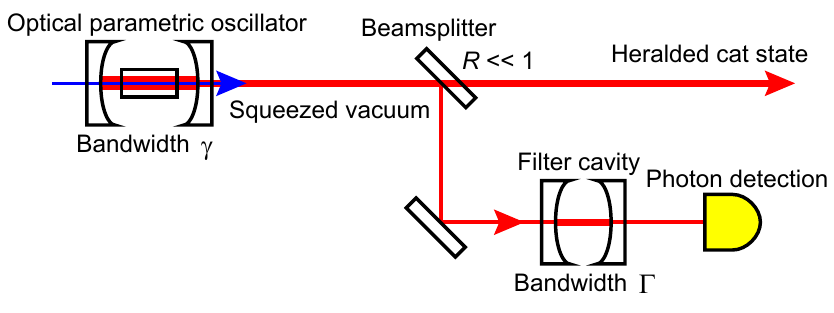}
\caption{
(Color online) 
Optical setup of our method inserting a filter cavity in the photon-subtraction path. 
Unlike conventional methods, by taking the bandwidth of the filter $\Gamma$ narrower than that of the OPO $\gamma$, the resulting photon-subtracted squeezed states become purer. 
}
\label{fig:Schematic}
\end{figure}

Here we theoretically show that the above mechanism indeed causes some inefficiency of heralded cat states in the ordinary CW methods. 
Furthermore, we also show that this inherent inefficiency can be arbitrarily suppressed by inserting a filter to extract a flat region of the spectrum before the photon detection. 
That is, here we propose a method potentially reaches to ideal photon-subtracted squeezed states in the CW regime, which are not obtainable with conventional methods. 
The schematic optical setup of our method is shown in Fig.~\ref{fig:Schematic}. 
Note that previous demonstrations of photon subtraction with CW methods are already using filter cavities, but their bandwidths are wider than those of OPOs, in order to utilize the raw correlations of photons produced by the OPO cavities as the wavepackets of heralded cat states. 
However, our calculations show that it is more advantageous to engineer the wavepackets of heralded cat states by filtering the subtraction path. 
As a side effect, the filtering by a cavity deforms the longitudinal mode function of the heralded cat states to an exponentially rising function, which is advantageous for real-time homodyne measurements \cite{Ogawa.prl2016}. 
Here we only discuss the case of one-photon subtraction, but the same mechanism works also for multiphoton subtraction.

In Sec.~\ref{sec:Basics}, we briefly summarize basic equations of photon subtraction in a single-mode regime. 
In Sec.~\ref{sec:Beam}, we deal with squeezed states of a beam (longitudinally infinite-mode states), with a general two-photon correlation, and photon subtraction from them. 
Impure squeezing of a wavepacket is discussed, by introducing the \textit{mode-match} of squeezing. 
Sec.~\ref{sec:OPO}, we apply our theory to the specific correlation created by a typical OPO. 
In Sec.~\ref{sec:Filter}, we show the impurity is arbitrarily suppressed by a filter cavity in the photon subtraction path. 
In Sec.~\ref{sec:summary}, we summarize the paper, and make additional observations.

\section{Basics of photon subtraction}
\label{sec:Basics}

In this section, we summarize basic descriptions of one-photon subtraction in the single-mode regime. 
Photon subtraction, which is application of the annihilation operator $\hat{a}$ to a quantum state, is approximated by beamsplitter tapping followed by a photon detection, as we show here.

We suppose the initial pure single-mode state to be $\ket{\psi}=\sum_{n=0}^\infty c_{n}\ket{n}$ which is expanded with respect to the photon-number eigenstates $\ket{n}\coloneqq(1/\sqrt{n})\hat{a}^{\dagger n}\ket{0}$. 
Here, $\hat{a}^\dagger$ is the creation operator, satisfying $[\hat{a},\hat{a}^\dagger] = 1$. 
The beamsplitter unitary operator is denoted by $\hat{B}(R)$, whose reflectivity $R$ is assumed to be small for the purpose of tapping. 
Expressing the ancillary tapping mode with the subscript `anc', 
\begin{align}
& \hat{B}(R)\ket{\psi}\otimes\ket{0}_\text{anc} \notag\\
= & \sum_{n=0}^\infty\frac{c_{n}[\hat{B}(R)\hat{a}^\dagger\hat{B}^\dagger(R)]^n}{\sqrt{n!}}\hat{B}(R)\ket{0}\otimes\ket{0}_\text{anc} \notag\\
= & \sum_{n=0}^\infty\frac{c_{n}(\sqrt{1-R}\hat{a}^\dagger+\sqrt{R}\hat{a}_\text{anc}^\dagger)^n}{\sqrt{n!}}\ket{0}\otimes\ket{0}_\text{anc} \notag\\
= & \sum_{n=0}^\infty c_{n}(\sqrt{1-R})^{n}\ket{n}\otimes\ket{0}_\text{anc} \notag\\
& + \sqrt{R}\sum_{n=1}^\infty c_{n}(\sqrt{1-R})^{n-1}\sqrt{n}\ket{n-1}\otimes\ket{1}_\text{anc} \notag\\
& + O(R). 
\label{eq:Subtraction}
\end{align}
Here, the invariance of vacuum states under beamsplitter operations $\hat{B}(R)\ket{0}\otimes\ket{0}_\text{anc}=\ket{0}\otimes\ket{0}_\text{anc}$ is used. 
Neglecting the noiseless attenuation terms $(\sqrt{1-R})^{n}$ \cite{Micuda.PRL2012}, which can arbitrarily approach the identity at the limit of a small reflectivity $R\to0$, the conditional state heralded by one-photon detection is, 
\begin{align}
_\text{anc}\bra{1}\hat{B}(R\to0)\ket{\psi}\otimes\ket{0}_\text{anc}
= & \sqrt{R}\sum_{n=1}^\infty c_{n}\sqrt{n}\ket{n-1} \notag\\
\propto & \hat{a}\ket{\psi}. 
\label{eq:SubtractionLimit}
\end{align}
Therefore, the above procedure mathematically approaches to the ideal photon subtraction which is a photon-annihilation operation. 
Note that the conditional success is quantum-mechanically inevitable because of the nonunitarity of the annihilation operator. 
However, wait-until-success methods, implemented with some quantum memories, can overcome this probabilistic nature.

Cat-state generation is achieved by applying this photon-subtraction process to a squeezed vacuum state. 
Single-mode squeezing operator with a squeezing parameter $\lvert r\rvert$ is defined as, 
\begin{align}
\hat{S}(r) = \exp\Bigl[\frac{1}{2}(r\hat{a}^\dagger\hat{a}^\dagger-r^\ast\hat{a}\hat{a})\Bigr], 
\end{align} 
where the superscript $\ast$ denotes the complex conjugation. 
The Bogoliubov transformation by the squeezing operator is $\hat{S}^\dagger(r)\hat{a}\hat{S}(r)=\hat{a}\cosh\lvert r\rvert + \hat{a}^\dagger\exp(2i\theta)\sinh\lvert r\rvert$, where $\theta\coloneqq\arg(r)/2$ expresses the antisqueezing direction. 
It is worth noting that a photon-subtracted squeezed state is equivalent to a squeezed single-photon state, from the following relation, 
\begin{align}
\hat{a}\hat{S}(r)\ket{0} 
= & \hat{S}(r)\hat{S}^\dagger(r)\hat{a}\hat{S}(r)\ket{0} \notag\\
= & \hat{S}(r)\bigl[\hat{a}\cosh\lvert r\rvert + \hat{a}^\dagger\exp(2i\theta)\sinh\lvert r\rvert\bigr]\ket{0} \notag\\
\propto & \hat{S}(r)\hat{a}^\dagger\ket{0}. 
\end{align}
This squeezed single-photon state well approximates a minus cat state with a small amplitude \cite{Lund.PRA2004}.

However, in real situations, we often have to deal with multimode quantum states, with some entanglement among modes. 
In the following sections, we deal with a squeezed vacuum state in a beam with time-translation symmetry (in a rotating frame), which is thus essentially multimode in the longitudinal direction.

\section{Impure squeezing of wavepacket}
\label{sec:Beam}

\subsection{Definition of a wavepacket}

From here, we deal with a light beam with the longitudinal coordinate $t$. 
The infinite-mode vacuum state of a beam $\ket{\emptyset}$, distinguished from a single-mode vacuum state $\ket{0}$, satisfies $\hat{a}(t)\ket{\emptyset}=0$ for all $t$. 
The instantaneous creation and annihilation operators $\hat{a}^\dagger(t)$ and $\hat{a}(t)$ satisfies $[\hat{a}(t),\hat{a}^\dagger(t^\prime)]=\delta(t-t^\prime)$, where $\delta(t)$ is the Dirac delta function. 
They have the Fourier counterparts, 
\begin{subequations}
\begin{align}
\hat{\widetilde{a}}^\dagger(\omega) = & \frac{1}{\sqrt{2\pi}}\int\hat{a}^\dagger(t)\exp(-i\omega t)d\omega, \\
\hat{\widetilde{a}}(\omega) = & \frac{1}{\sqrt{2\pi}}\int\hat{a}(t)\exp(i\omega t)d\omega, 
\end{align}
\end{subequations}
with $[\hat{\widetilde{a}}(\omega),\hat{\widetilde{a}}^\dagger(\omega^\prime)]=\delta(\omega-\omega^\prime)$. 
We take a rotating frame so that the degenerate frequency of squeezing becomes $\omega=0$.

Quadrature operators with respect to a phase $\theta$ are defined as, 
\begin{align}
\hat{\widetilde{x}}^{(\theta)}(\omega) \coloneqq & \frac{\hat{\widetilde{a}}(\omega)e^{-i\theta}+\hat{\widetilde{a}}^\dagger(\omega)e^{i\theta}}{\sqrt{2}} \notag\\
\hat{\widetilde{p}}^{(\theta)}(\omega) \coloneqq & \frac{\hat{\widetilde{a}}(\omega)e^{-i\theta}-\hat{\widetilde{a}}^\dagger(\omega)e^{i\theta}}{\sqrt{2}i}, 
\end{align}
and the same definition applies to the time domain, $\hat{x}^{(\theta)}(t)$ and $\hat{p}^{(\theta)}(t)$. 
We omit the superscript phase $(\theta)$ when $\theta=0$.

We define a creation operator and an annihilation operator associated with a wavepacket mode function $g(t)$ as, 
\begin{subequations}
\begin{align}
\hat{a}_g^\dagger \coloneqq & \int g(t)\hat{a}^\dagger(t)dt = \int \widetilde{g}(\omega)\hat{\widetilde{a}}^\dagger(\omega)d\omega, \\
\hat{a}_g \coloneqq & \int g^\ast(t)\hat{a}(t)dt = \int \widetilde{g}^\ast(\omega)\hat{\widetilde{a}}(\omega)d\omega, 
\end{align}
\end{subequations}
where the Fourier pair of a complex function is defined as, 
\begin{align}
\widetilde{f}(\omega) \coloneqq \frac{1}{\sqrt{2\pi}}\int f(t)\exp(i\omega t)dt. 
\end{align}
The bosonic commutation relation $[\hat{a}_{g_k},\hat{a}_{g_\ell}^\dagger] = \delta_{k\ell}$ with the Kronecker delta $\delta_{k\ell}$ is equivalent to the orthonormalization condition of the wavepacket mode functions, 
\begin{align}
\langle g_k,g_\ell\rangle 
\coloneqq & \int g_k^\ast(t)g_\ell(t)dt 
= \int \widetilde{g}_k^\ast(\omega)\widetilde{g}_\ell(\omega)d\omega \notag\\
= & \delta_{k\ell}. 
\end{align}
Here, the inner product between two functions $f(t)$ and $f^\prime(t)$ is expressed by $\langle f,f^\prime\rangle$. 
The norm of a function $f(t)$ is defined as $\lVert f \rVert \coloneqq \sqrt{\langle f,f\rangle}$. 
We express normalization of a function $f(t)$ or $\widetilde{f}(\omega)$ as, 
\begin{align}
N(f)(t) & \coloneqq \frac{f(t)}{\lVert f \rVert}, &
N(\widetilde{f})(\omega) & \coloneqq \frac{\widetilde{f}(\omega)}{\lVert f \rVert}. 
\end{align} 
We assume the normalization for the wavepacket mode function $g(t) = N(g)(t)$.

The corresponding quadrature operators are $\hat{x}_g \coloneqq [\hat{a}_g+\hat{a}_g^\dagger]/\sqrt{2}$ and $\hat{p}_g \coloneqq [\hat{a}_g-\hat{a}_g^\dagger]/(\sqrt{2}i)$. 
In particular, if we take $g(t)$ as a real function (remember that we are in the rotating frame), thus $\widetilde{g}(-\omega) = \widetilde{g}^\ast(\omega)$, in this case, $\hat{x}_g=\int g(t)\hat{x}(t)dt$ and $\hat{p}_g=\int g(t)\hat{p}(t)dt$.

\subsection{Continuous squeezing operator}

In general, a unitary squeezing operator on a beam can be expressed in the form of 
\begin{align}
\hat{S}_r = \exp\Bigl[\frac{1}{2}(\hat{P}_r^\dagger - \hat{P}_r)\Bigr], 
\label{eq:Squeezing}
\end{align}
with photon-pair creation and annihilation operators $\hat{P}_r^\dagger$ and $\hat{P}_r$, defined by using a photon-pair correlation function $r(t_1,t_2)$ as, 
\begin{align}
\hat{P}_r^\dagger 
\coloneqq & \iint r(t_1,t_2)\hat{a}^\dagger(t_1)\hat{a}^\dagger(t_2)dt_1dt_2. 
\end{align}
The continuous pumping appears as time-translation symmetry, $r(t_1,t_2)=r(t_1-t_2)$, in the rotating frame. 
Under this time-translation symmetry, the upper and lower sidebands at each frequency are exclusively coupled by the squeezing operator, as 
\begin{align}
\hat{P}_r^\dagger 
= & \iint r(t_1-t_2)\hat{a}^\dagger(t_1)\hat{a}^\dagger(t_2)dt_1dt_2 \notag\\
= & \iint \widetilde{r}(\omega)\hat{\widetilde{a}}^\dagger(\omega)\hat{\widetilde{a}}^\dagger(-\omega)d\omega. 
\label{eq:PairCreation}
\end{align}
From the symmetry between $\hat{a}^\dagger(t_1)$ and $\hat{a}^\dagger(t_2)$, the correlation function has the time-reversal symmetry $r(t)=r(-t)$, which leads to $\widetilde{r}(\omega)=\widetilde{r}(-\omega)$. 
The squeezing operator makes a Bogoliubov transformation, 
\begin{align}
\hat{S}_r^\dagger\hat{\widetilde{a}}(\omega)\hat{S}_r 
= & \hat{\widetilde{a}}(\omega)\cosh \lvert\widetilde{r}(\omega)\rvert \notag\\
& + \hat{\widetilde{a}}^\dagger(-\omega) \exp[2i\theta(\omega)] \sinh \lvert\widetilde{r}(\omega)\rvert, 
\label{eq:BogoliubovCW}
\end{align}
with $\theta(\omega) \coloneqq \arg[\widetilde{r}(\omega)]/2$, which expresses pure squeezing in each sideband pair, 
\begin{subequations}
\begin{align}
& \hat{S}_r^\dagger\bigl[\hat{\widetilde{x}}^{(\theta(\omega))}(\omega)+\hat{\widetilde{x}}^{(\theta(\omega))}(-\omega)\bigr]\hat{S}_r \notag\\
& = \bigl[\hat{\widetilde{x}}^{(\theta(\omega))}(\omega)+\hat{\widetilde{x}}^{(\theta(\omega))}(-\omega)\bigr]\exp[\lvert\widetilde{r}(\omega)\rvert], \\
& \hat{S}_r^\dagger\bigl[\hat{\widetilde{p}}^{(\theta(\omega))}(\omega)+\hat{\widetilde{p}}^{(\theta(\omega))}(-\omega)\bigr]\hat{S}_r \notag\\
& = \bigl[\hat{\widetilde{p}}^{(\theta(\omega))}(\omega)+\hat{\widetilde{p}}^{(\theta(\omega))}(-\omega)\bigr]\exp[-\lvert\widetilde{r}(\omega)\rvert].
\end{align}
\end{subequations}
Conversely, given an antisqueezing spectrum $V^{(+)}(\omega)$ and a squeezing spectrum $V^{(-)}(\omega)$ which are pure, $V^{(+)}(\omega)=1/V^{(-)}(\omega)=\exp(2 \lvert\widetilde{r}(\omega)\rvert)$, and given a squeezing phase $\theta(\omega)$ at each frequency, the squeezing operator $\hat{S}_r$ is expressed uniquely in the form of eqs.~\eqref{eq:Squeezing} and \eqref{eq:PairCreation}.

\subsection{Photon subtraction}

We suppose the timing of the photon subtraction to be $t=0$ without loss of generality. 
The whole state after the photon subtraction is, by using the Bogoliubov transformation in eq.~\eqref{eq:BogoliubovCW}, the time-reversal symmetry $\widetilde{r}(\omega)=\widetilde{r}(-\omega)$, and $\hat{\widetilde{a}}(\omega)\ket{\emptyset}=0$, 
\begin{align}
& \hat{a}(0)\hat{S}_r\ket{\emptyset} \notag\\
= & \biggl[\int\hat{\widetilde{a}}(\omega)d\omega\biggr]\hat{S}_r\ket{\emptyset} \notag\\
= & \hat{S}_r\biggl\{\int\hat{\widetilde{a}}^\dagger(\omega)\exp[2i\theta(\omega)]\sinh \lvert\widetilde{r}(\omega)\rvert d\omega\biggr\}\ket{\emptyset}. 
\label{eq:SqueezedSinglePhotonCW}
\end{align}
In particular, when the squeezing is weak, $\sinh \lvert\widetilde{r}(\omega)\rvert \approx \lvert\widetilde{r}(\omega)\rvert$, and the state approaches to, 
\begin{align}
\hat{a}(0)\hat{S}_r\ket{\emptyset} 
\approx \hat{S}_r\biggl[\int \widetilde{r}(\omega)\hat{\widetilde{a}}^\dagger(\omega) d\omega\biggr] \ket{\emptyset} 
\propto \hat{S}_r\hat{a}_{N(r)}^\dagger \ket{\emptyset}. 
\label{eq:GeneralWavepacket}
\end{align}
Therefore, the heralded cat state is equivalent to a single photon state in a wavepacket mode close to $N(r)(t)$ exposed to longitudinally multimode squeezing $\hat{S}_r$. 
The deviation of the wavepacket mode from $N(r)(t)$ has order of $O(\lvert\widetilde{r}(\omega)\rvert^3)$, which can be checked from the Taylor expansion of $\sinh \lvert\widetilde{r}(\omega)\rvert$. 
That is, the same order of contribution as five-photon terms.

We discuss below that squeezing with a nonflat spectrum operating on such a wavepacket induces some impurity \cite{Takeoka.PRA2008}, by making entanglement with orthogonal modes.

\subsection{Natural impurity in wavepacket squeezing}

If a continuous squeezing process involves no optical losses, the squeezing at each frequency can be pure, expressed by a Bogoliubov transformation in eq.~\eqref{eq:BogoliubovCW}. 
However, even in the case of pure sideband squeezing confirmed by the minimum uncertainty $V^{(+)}(\omega)V^{(-)}(\omega)=1$, this minimum uncertainty is not preserved in general for the squeezing with respect to a wavepacket $g(t)$. 
In order to discuss this, for simplicity we consider the case the wavepacket mode function $g(t)$ is a real function. 
Furthermore, we suppose $\widetilde{r}(\omega)\in\mathbb{R}^+$ for all $\omega$, and under this condition $\hat{\widetilde{x}}(\omega)+\hat{\widetilde{x}}(-\omega)$ quadrature is antisqueezed and $\hat{\widetilde{p}}(\omega)+\hat{\widetilde{p}}(-\omega)$ quadrature is squeezed. 
The quadrature variance of the wavepacket mode $g(t)$ is, 
\begin{subequations}
\begin{align}
& \bra{\emptyset}\bigl(\hat{S}_r^\dagger\hat{x}_g\hat{S}_r\bigr)^2\ket{\emptyset} \notag\\
& = \frac{1}{2}\int \lvert \widetilde{g}(\omega) \rvert^2 V^{(+)}(\omega) d\omega 
\coloneqq \frac{1}{2}V_g^{(+)}, \\
& \bra{\emptyset}\bigl(\hat{S}_r^\dagger\hat{p}_g\hat{S}_r\bigr)^2\ket{\emptyset} \notag\\
& = \frac{1}{2}\int \lvert \widetilde{g}(\omega) \rvert^2 V^{(-)}(\omega) d\omega 
\coloneqq \frac{1}{2}V_g^{(-)}. 
\end{align}\label{eq:SqueezingWavepacket}
\end{subequations}
Note that $\int\lvert\widetilde{g}(\omega)\rvert^2d\omega = \lVert g \rVert^2 = 1$, and that $1/2$ is the quadrature variance of a vacuum state. 
Apparently, $V_g^{(+)}V_g^{(-)}>1$ unless the spectra in the relevant domain are flat. 
This can be shown by using the Cauchy-Schwarz inequality, 
\begin{align}
& \int \lvert \widetilde{g}(\omega) \rvert^2 V^{(+)}(\omega) d\omega
\int \lvert \widetilde{g}(\omega^\prime) \rvert^2 V^{(-)}(\omega^\prime) d\omega^\prime \notag\\
& \ge \biggl\lvert \int \lvert\widetilde{g}(\omega)\rvert^2 \sqrt{V^{(+)}(\omega)V^{(-)}(\omega)} d\omega \biggr\rvert^2 \ge \lVert g \rVert^4, 
\end{align}
where the first inequality becomes the equality if and only if $\widetilde{g}(\omega)\sqrt{V^{(+)}(\omega)} \propto \widetilde{g}(\omega)\sqrt{V^{(-)}(\omega)}$ which is possible for the pure sideband squeezing only when the spectra are flat for nonzero $\widetilde{g}(\omega)$.

\subsection{Decomposition of a photon-pair operator and mode-match}
\label{ssec:Decomposition}

In order to understand the impurity, here we introduce the \textit{mode-match} of squeezing, associated with a pair-creation operator in eq.~\eqref{eq:PairCreation}, with a wavepacket mode $g(t)$. 
We use the following relation, 
\begin{align}
[\hat{a}_g,\hat{P}_r^\dagger] 
= & 2\int g^\ast(\tau)r(t-\tau)\hat{a}^\dagger(t) d\tau dt \notag\\
= & 2\int (g^\ast \ast r)(t)\hat{a}^\dagger(t) d\tau dt \notag\\
= & 2\lVert g^\ast \ast r \rVert \hat{a}_{N(g^\ast \ast r)}^\dagger, 
\end{align}
where $(f\ast f^\prime)(t)$ denotes the convolution of $f(t)$ and $f^\prime(t)$. 
Here, the coefficient $2$ comes from the symmetry between $\hat{a}^\dagger(t_1)$ and $\hat{a}^\dagger(t_2)$. 
Furthermore, we can decompose $N(g^\ast \ast r)(t)$ into a portion along $g(t)$ and a portion orthogonal to it via the Gram-Schmidt orthogonalization, 
\begin{align}
N(g^\ast \ast r)(t) 
= & \langle g, N(g^\ast \ast r)\rangle g(t) \notag\\
& + \sqrt{1-\lvert\langle g, N(g^\ast \ast r)\rangle\rvert^2} g^\perp(t), 
\end{align}
where $g^\perp(t)$ is a normalized function satisfying $\langle g,g^\perp\rangle=0$. 
From above, we have derived, 
\begin{align}
\hat{P}_r^\dagger 
= & \langle g, g^\ast \ast r\rangle\hat{a}_g^{\dagger2} + 2\sqrt{\lVert g^\ast \ast r \rVert^2-\lvert\langle g, g^\ast \ast r\rangle\rvert^2}\hat{a}_g^\dagger\hat{a}_{g^\perp}^\dagger \notag\\
& + (\text{other irrelevant terms}). 
\label{eq:SpecificWavepacketPair}
\end{align}
Here, ``irrelevant terms'' means that they commute with $\hat{a}_g$, but does not mean that they commute with ${a}_{g^\perp}$. 
In fact, taking $g_0(t)=g(t)$ and $g_1(t)=g^\perp(t)$, and repeating the Gram-Schmidt orthogonalization of $(g_k^\ast \ast r)(t)$ to obtain $g_{k+1}(t)$, we reach the form, 
\begin{align}
\hat{P}_r^\dagger = \sum_k \bigl(c_{k,k}\hat{a}_{g_k}^{\dagger2} + 2c_{k,k+1}\hat{a}_{g_k}^\dagger \hat{a}_{g_{k+1}}^\dagger\bigr)+\hat{P}_{r\setminus[g]}^\dagger, 
\label{eq:WavepacketSeries}
\end{align}
with a set of orthonormal functions $\{g_k(t)\}_{k\in\mathbb{N}}$ and a set of coefficients $\{c_{k,k},c_{k,k+1}\}_{k\in\mathbb{N}}$, where $\hat{P}_{r\setminus[g]}^\dagger$ expresses the remaining part of the pair-creation operator in modes totally disconnected from $g(t)$, if there is.

Therefore, if $N(g^\ast \ast r)(t)$ is close to $g(t)$ (up to the global phase), $\hat{a}_g^{\dagger2}$ terms are dominant compared with $2\hat{a}_g^\dagger\hat{a}_{g^\perp}^\dagger$ terms, which means the squeezing is almost pure. 
Otherwise, $2\hat{a}_g^\dagger\hat{a}_{g^\perp}^\dagger$ terms become innegligible, and photons are not always generated in pairs in the wavepacket $g(t)$ but some pairs make entanglement with an orthogonal mode. 
The ratio between squared coefficients (corresponding to probability) of $\hat{a}_g^{\dagger2}$ (single-mode squeezing) and $2\hat{a}_g^\dagger\hat{a}_{g^\perp}^\dagger$ (two-mode squeezing) is $\lvert c_{g,g}\rvert^2:\lvert c_{g,g^\perp}\rvert^2=\lvert\langle g, N(g^\ast \ast r)\rangle\rvert^2:[1-\lvert\langle g, N(g^\ast \ast r)\rangle\rvert^2]$, and thus, in this sense, we define, 
\begin{align}
M[g,r] \coloneqq \lvert\langle N(g), N(g^\ast \ast r)\rangle\rvert^2, 
\label{eq:ModeMatch}
\end{align}
as the \textit{mode-matching rate} of the squeezing $\hat{S}_r$ with respect to the wavepacket mode $N(g)(t)$. 
(Here, the assumption of normalized $g(t)$ is forgotten and the normalization is explicitly included in the definition for later convenience.)

The above mode-matching condition, $M[g,r]$ being closer to $1$, is decomposed into two parts: first, $g(t)\approx e^{i\varphi}g^\ast(t)$ with a fixed phase $\varphi$, and second, the convolution with $r(t)$ does not largely deform $g(t)$. 
Note that a sufficient condition of the first condition is the wavepacket function $g(t)$ being a real function. 
The first condition means that $g(t)$ takes the upper and lower sidebands symmetrically, which is checked by considering the relation with the Fourier counterpart $\widetilde{g}(\omega)$. 
This is related to the energy conservation law among pump, signal and idler photons. 
From the second condition, we can expect that a narrower bandwidth of the wavepacket $g(t)$ is more advantageous to increase the mode-match.

\section{Optical parametric oscillator}
\label{sec:OPO}

\subsection{Sideband squeezing}

Now we apply the above discussion to an ideal OPO squeezing. 
We consider the ideal case where a beam of a squeezed vacuum state from an OPO never suffers from any optical losses. 
Referring to previous works \cite{Collet.PRA1984,Takeoka.PRA2008}, the ideal OPO squeezing is a unitary Bogoliubov transformation, 
\begin{align}
& \hat{S}_{\gamma,\epsilon}^\dagger\hat{\widetilde{a}}(\omega)\hat{S}_{\gamma,\epsilon} \notag\\
= &  \frac{\gamma^2+\omega^2+\lvert\epsilon\rvert^2}{(\gamma-i\omega)^2-\lvert\epsilon\rvert^2}\hat{\widetilde{a}}(\omega) + \frac{2\gamma\epsilon}{(\gamma-i\omega)^2-\lvert\epsilon\rvert^2}\hat{\widetilde{a}}^\dagger(-\omega), 
\label{eq:BogoliubovOPO}
\end{align}
where $\gamma\in\mathbb{R}^+$ is the cavity decay constant (with the factor $2$), and $\epsilon\in\mathbb{C}$ represents the pump field. 
We only consider $\lvert\epsilon\rvert<\gamma$, and the maximum squeezing is obtained at $\omega=0$ at the limit of the oscillation threshold $\lvert\epsilon\rvert\to\gamma$.

We here suppose $\epsilon\in\mathbb{R}^+$ for simplicity without loss of generality, by which $\hat{\widetilde{p}}(0)$ quadrature is squeezed. 
Eq.~\eqref{eq:BogoliubovOPO} is equivalent to, 
\begin{align}
& \hat{S}_{\gamma,\epsilon}^\dagger[\hat{\widetilde{a}}(\omega)\pm\hat{\widetilde{a}}^\dagger(-\omega)]\hat{S}_{\gamma,\epsilon} \notag\\
= & \frac{\gamma\pm\epsilon+i\omega}{\gamma\mp\epsilon-i\omega}[\hat{\widetilde{a}}(\omega)\pm\hat{\widetilde{a}}^\dagger(-\omega)], 
\end{align}
and from this, we obtain the sideband antisqueezing and squeezing, 
\begin{subequations}
\begin{align}
& \hat{S}_{\gamma,\epsilon}^\dagger[\hat{\widetilde{x}}(\omega)+\hat{\widetilde{x}}(-\omega)]\hat{S}_{\gamma,\epsilon} \notag\\
= & \Bigl\lvert\frac{\gamma+\epsilon+i\omega}{\gamma-\epsilon-i\omega}\Bigr\rvert\bigl[\hat{\widetilde{x}}^{(-\phi(\omega))}(\omega)+\hat{\widetilde{x}}^{(-\phi(-\omega))}(-\omega)\bigr], \\
& \hat{S}_{\gamma,\epsilon}^\dagger[\hat{\widetilde{p}}(\omega)+\hat{\widetilde{p}}(-\omega)]\hat{S}_{\gamma,\epsilon} \notag\\
= & \Bigl\lvert\frac{\gamma-\epsilon+i\omega}{\gamma+\epsilon-i\omega}\Bigr\rvert\bigl[\hat{\widetilde{p}}^{(-\phi(\omega))}(\omega)+\hat{\widetilde{p}}^{(-\phi(-\omega))}(-\omega)\bigr], 
\end{align}
\end{subequations}
with the phase rotation, 
\begin{align}
\phi(\omega) 
= \arg\Bigl(\frac{\gamma+\epsilon+i\omega}{\gamma-\epsilon-i\omega}\Bigr)
= \arg\Bigl(\frac{\gamma-\epsilon+i\omega}{\gamma+\epsilon-i\omega}\Bigr), 
\end{align}
which remains to be nonzero at $\epsilon\to0$ for $\omega\neq0$ and thus is coming from the cavity, expressing the time delay by storage. 
This phase rotation by $\phi(\omega)$ can be neglected when we consider a squeezed vacuum output $\hat{S}_{\gamma,\epsilon}\ket{\emptyset}$, because the input vacuum state $\ket{\emptyset}$ is rotation-invariant. 
The antisqueezing spectrum $V_{\gamma,\epsilon}^{(+)}$ and the squeezing spectrum $V_{\gamma,\epsilon}^{(-)}$ are, 
\begin{subequations}
\label{eq:SqueezingSpectrum}
\begin{align}
V_{\gamma,\epsilon}^{(+)}(\omega) 
= & \Bigl\lvert\frac{\gamma+\epsilon+i\omega}{\gamma-\epsilon-i\omega}\Bigr\rvert^2 
= \frac{(\gamma+\epsilon)^2+\omega^2}{(\gamma-\epsilon)^2+\omega^2}, \\
V_{\gamma,\epsilon}^{(-)}(\omega) 
= & \Bigl\lvert\frac{\gamma-\epsilon+i\omega}{\gamma+\epsilon-i\omega}\Bigr\rvert^2 
= \frac{(\gamma-\epsilon)^2+\omega^2}{(\gamma+\epsilon)^2+\omega^2}. 
\end{align}
\end{subequations}
The squeezing at each frequency $\omega$ is pure with minimum uncertainty, i.e., $V_{\gamma,\epsilon}^{(+)}(\omega)V_{\gamma,\epsilon}^{(-)}(\omega)=1$ for all $\omega$, which is a natural result from the assumption of no optical losses and continuous pumping.

\subsection{Equivalent squeezing operator}

From above, we may redefine the squeezing operator of the ideal OPO as, 
\begin{align}
\hat{S}_{\gamma,\epsilon} 
= & \exp\Bigl\{\frac{1}{2}\int\bigl[\widetilde{r}_{\gamma,\epsilon}(\omega)\hat{\widetilde{a}}^\dagger(\omega)\hat{\widetilde{a}}^\dagger(-\omega) \notag\\
& \quad - \widetilde{r}_{\gamma,\epsilon}(\omega)\hat{\widetilde{a}}(\omega)\hat{\widetilde{a}}(-\omega)\bigr]d\omega\Bigr\}, 
\end{align}
with the frequency-dependent squeezing parameter, 
\begin{align}
\widetilde{r}_{\gamma,\epsilon}(\omega) = \ln \sqrt{V_{\gamma,\epsilon}^{(+)}(\omega)}
= \frac{1}{2}\ln\biggl[\frac{(\gamma+\epsilon)^2+\omega^2}{(\gamma-\epsilon)^2+\omega^2}\biggr], 
\label{eq:SqueezingParameter}
\end{align}
by neglecting the phase rotation by $\phi(\omega)$. 
We will use this definition in the following. 
The squeezing operator has the equivalent time-domain representation, 
\begin{align}
\hat{S}_{\gamma,\epsilon} 
= & \exp\Bigl\{\frac{1}{2}\int\bigl[r_{\gamma,\epsilon}(t_1-t_2)\hat{a}^\dagger(t_1)\hat{a}^\dagger(t_2) \notag\\
& \quad - r_{\gamma,\epsilon}(t_1-t_2)\hat{a}(t_1)\hat{a}(t_2)\bigr]dt_1dt_2\Bigr\}. 
\end{align}

\begin{figure}[tb]
\centering
\includegraphics{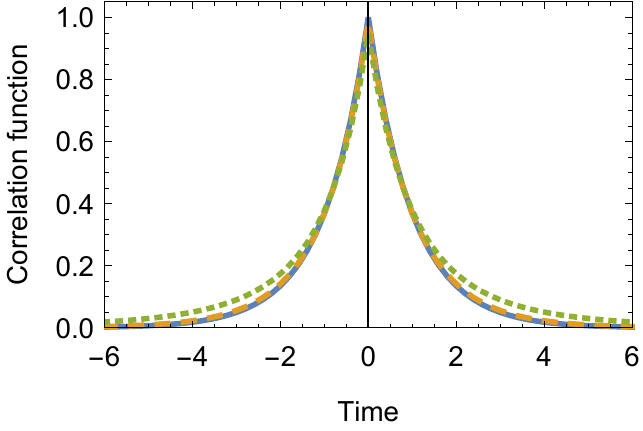}
\caption{
(Color online) 
Normalized correlation functions $N(r_{\gamma,\epsilon})(t)$ in the time domain, obtained by numerical calculations, for $\gamma = 1$. 
Solid blue: both-side exponential function corresponding to the limit of $\epsilon\to0$. 
Dashed orange: $\epsilon=0.3$. 
Dotted green: $\epsilon=0.7$. 
The latter two correspond to the squeezing degrees of about $5.4$ dB and $15.1$ dB, respectively, at $\omega=0$. 
}
\label{fig:Correlation}
\end{figure}

An important point is, when the pump field is weak ($\epsilon/\gamma\ll1$), 
\begin{subequations}
\label{eq:CorrWeakOPO}
\begin{align}
\widetilde{r}_{\gamma,\epsilon}(\omega) 
& \approx \epsilon \frac{2\gamma}{\gamma^2+\omega^2}, \\
r_{\gamma,\epsilon}(t) 
& \approx \epsilon\sqrt{2\pi}\exp(-\gamma\lvert t\rvert). \label{eq:BothSideExponential}
\end{align}
\end{subequations}
This both-side exponential function is a typical wavepacket mode function of a cat state \cite{Molmer.PRA2006}, from the relation in eq.~\eqref{eq:GeneralWavepacket}. 
As shown in Fig.~\ref{fig:Correlation}, the normalized correlation function $N(r_{\gamma,\epsilon})(t)$ is very close to the both-side exponential function unless the squeezing level is very high.

\subsection{Explanation of both-side exponential correlation}

The both-side exponential function in eq.~\eqref{eq:CorrWeakOPO} is understood as follows. 
We may write the correlation function of a general pair-creation operator in eq.~\eqref{eq:PairCreation} as, 
\begin{align}
r(t_1-t_2) = \int \lambda(t_1-\tau)\lambda(t_2-\tau)d\tau, 
\end{align}
or shortly $r = \lambda \ast \lambda^R$, where the superscript $R$ denotes the time reversal $f^R(t) \coloneqq f(-t)$. 
Furthermore, we may interpret $\tau$ as the timing of a pump photon to be converted to a photon pair inside the cavity, and take $\lambda$ as the cavity decay function, 
\begin{align}
\lambda(t) \propto \exp(-\gamma t)u(t), 
\end{align}
where $u(t)$ is the unit step function, 
\begin{align}
u(t) = 
\begin{cases}
0, & t<0, \\
1, & t\ge0. 
\end{cases}
\end{align}
In this case, the correlation function $r(t)$ becomes a both-side exponential function, 
\begin{align}
r(t) \propto \exp(-\gamma\lvert t\rvert), 
\end{align}

On the other hand, an interesting thing is that, when the parametric down conversion is not degenerate (e.g. in polarization), the decay of signal and idler photons can be asymmetric, $r=\lambda_\text{sig}\ast\lambda_\text{idl}^R$ with $\lambda_\text{sig}\neq\lambda_\text{idl}$. 
In this case, the time-reversal symmetry of the correlation function is broken, $r\neq r^R$. 
This mechanism is found to be useful, e.g., in creation of exponentially rising wavepackets $g(t)\propto\exp(\gamma t)u(-t)$ of heralded single-photon states by setting $\lambda_\text{sig}(t) \propto \delta(t)$, which is advantageous in real-time homodyne measurements \cite{Ogawa.prl2016}.

\begin{figure}[tb]
\centering
\includegraphics{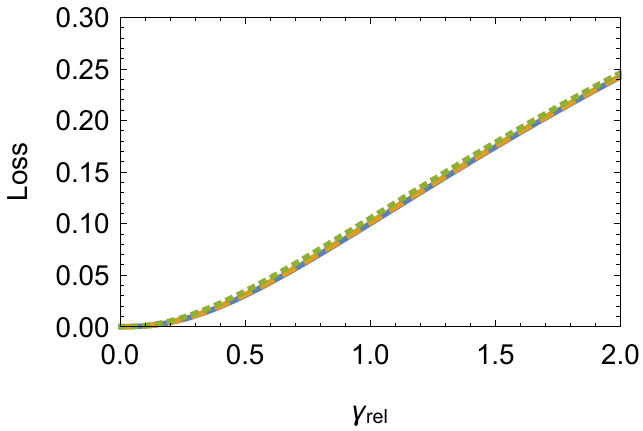}
\caption{
(Color online) 
Equivalent amount of losses $L$ as a function of the relative bandwidth of the wavepacket mode function $\gamma_\text{rel}$. 
Solid blue: $\epsilon/\gamma=0.03$. 
Dashed orange: $\epsilon/\gamma=0.3$. 
Dotted green: $\epsilon/\gamma=0.7$. 
They correspond to the squeezing degrees of about $0.5$ dB, $5.4$ dB, and $15.1$ dB, respectively, at $\omega=0$. 
The traces are almost overlapped. 
In particular, $L\approx0.1$ at $\gamma_\text{rel}=1$. 
}
\label{fig:EquivLoss}
\end{figure}

\subsection{Impurity and equivalent losses}

In order to estimate how the bandwidth of the wavepacket affect the impurity of squeezing, we calculate the squeezing and antisqueezing with respect to wavepackets, 
\begin{subequations}
\begin{align}
g(t) = & \sqrt{\gamma_\text{rel}\gamma}\exp(-\gamma_\text{rel}\gamma\lvert t\rvert), \label{eq:LorentzTD}\\
\widetilde{g}(\omega) = & \sqrt{\frac{2\gamma_\text{rel}\gamma}{\pi}}\frac{\gamma_\text{rel}\gamma}{\gamma_\text{rel}^2\gamma^2+\omega^2}. 
\end{align}
\end{subequations}
by using eq.~\eqref{eq:SqueezingWavepacket}. 
Here, $\gamma_\text{rel}$ denotes the relative bandwidth which is dimensionless, and $\gamma_\text{rel} = 1$ corresponds to the typical both-side exponential wavepacket function of heralded cat states, proportional to eq.~\eqref{eq:BothSideExponential}. 
Then, we quantify the asymmetry between the squeezing and the antisqueezing via the equivalent amount of losses $L$. 
Linear optical losses are equivalent to invasion of vacuum fluctuation from a virtual beamsplitter. 
When a pure single-mode squeezed state with a squeezing parameter $r>0$ suffers from losses $L$, the minimum uncertainty relation is broken, and the squeezed and antisqueezed quadrature variances $(1/2)V^{(+)}$ and $(1/2)V^{(-)}$ become, 
\begin{subequations}
\begin{align}
\frac{1}{2}V^{(+)} = & \frac{1}{2}\bigl[(1-L)e^{2r}+L\bigr], \\
\frac{1}{2}V^{(-)} = & \frac{1}{2}\bigl[(1-L)e^{-2r}+L\bigr], 
\end{align}
\end{subequations}
and therefore, the equivalent amount of losses $L$ to express the asymmetry is, 
\begin{align}
L = \frac{V^{(+)}V^{(-)}-1}{V^{(+)}+V^{(-)}-2}. 
\end{align}
Note that $L$ is indefinite when $V^{(+)}=V^{(-)}=1$, which corresponds to the fact that a vacuum state is not changed by losses. 
The calculated $L$ as a function of $\gamma_\text{rel}$ is plotted in Fig.~\ref{fig:EquivLoss}, which is slightly dependent on the degree of squeezing $\epsilon/\gamma$. 
The dependence on $\epsilon/\gamma$ is so small that we cannot almost see this dependence from the figure. 
At $\gamma_\text{rel}=1$, the corresponding amount of losses is about $10\%$, therefore we may consider the ordinary cat-generation methods include this $10\%$ of losses in some form, which will be discussed in Sec.~\ref{ssec:ImpurePhotonSubtract}.

\subsection{Mode-matching rate of squeezing}

Now we see that the $10\%$ of the equivalent losses coincides with the mode-matching rate discussed in Sec.~\ref{ssec:Decomposition}.

We assume the both-side exponential correlation function $r(t) = r^\ast(t) \propto \exp(-\gamma \lvert t\rvert)$ of the OPO squeezing. 
For the wavepacket with $\gamma_\text{rel} = 1$, $g(t)$ in eq.~\eqref{eq:LorentzTD} coincides with the normalized correlation function $N(r)(t)$. 
By using 
\begin{align}
N(r \ast r)(t) = \sqrt{\frac{2\gamma}{5}}\Bigl(1+\gamma\lvert t\rvert\Bigr)\exp(-\gamma\lvert t\rvert), 
\end{align}
the mode-matching rate in eq.~\eqref{eq:ModeMatch} is calculated as, 
\begin{align}
M[r,r] = \lvert\langle N(r), N(r \ast r)\rangle\rvert^2 
= \biggl\lvert \frac{3}{\sqrt{10}}\biggr\rvert^2=\frac{9}{10}. 
\label{eq:BareModeMatch}
\end{align}
This explains the equivalent losses being about $10\%$ at $\gamma_\text{rel} = 1$. 
Figure~\ref{fig:ModeFuncFiltered} shows the functions $N(r)(t)$ and $N(r\ast r)(t)$, and their discrepancy corresponds to the inefficiency of the squeezing.

\begin{figure}[tb]
\centering
\includegraphics{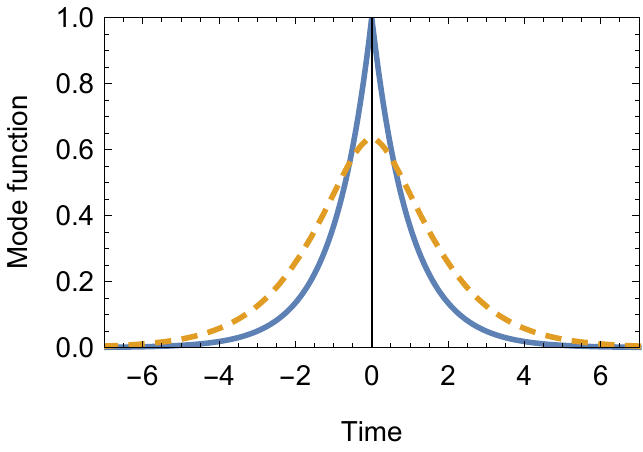}
\caption{
(Color online) 
Mode functions to show the mode-match of squeezing to heralded single-photon wavepackets, for the case without filtering, for $r(t)\propto\exp(-\gamma\lvert t\rvert)$ with $\gamma=1$. 
Solid blue: $N(r)(t)$. 
Dashed orange: $N(r\ast r)(t)$. 
}
\label{fig:ModeFuncBare}
\end{figure}

\subsection{Impurity in photon subtraction}
\label{ssec:ImpurePhotonSubtract}

A remarkable consideration is that optical losses commute with the photon subtraction process. 
When an annihilation operator $\hat{a}$ is applied to some single-mode state $\ket{\psi}$ which suffers from losses as $\hat{B}(L)\ket{\psi}\otimes\ket{0}_\text{anc}$, where $\hat{B}(L)$ is the beamsplitter interaction used in Sec.~\ref{sec:Basics}, then, it transforms as, 
\begin{align}
& \hat{a}\hat{B}(L)\ket{\psi}\otimes\ket{0}_\text{anc} \notag\\
= & \hat{B}(L)\bigl[\hat{B}^\dagger(1-L)\hat{a}\hat{B}(1-L)\bigr]\ket{\psi}\otimes\ket{0}_\text{anc} \notag\\
= & \hat{B}(L)(\sqrt{1-L}\hat{a}-\sqrt{L}\hat{a}_\text{anc})\ket{\psi}\otimes\ket{0}_\text{anc} \notag\\
\propto & \hat{B}(L)\hat{a}\ket{\psi}\otimes\ket{0}_\text{anc}. 
\label{eq:LossPreservation}
\end{align}
Therefore, the losses that the initial state $\ket{\psi}$ has suffered appear as the losses that the ideal photon-subtracted state $\hat{a}\ket{\psi}$ suffers. 
If this theory is applied to the virtual losses due to the nonflat squeezing spectrum discussed above, it leads to the conclusion that the ideal photon-subtracted state suffers from about $10\%$ of optical losses.

However, somewhat surprisingly, we can see that this is not the case, from eq.~\eqref{eq:GeneralWavepacket}. 
In the limit of weak pumping, the conditional state approaches to an ideal heralded single-photon state in the wavepacket $N(r_{\gamma,\epsilon})$, rather than that with $10\%$ of losses. 
The true situation is that an ideal heralded single-photon state $\hat{a}_{N(r_{\gamma,\epsilon})}^\dagger\ket{\emptyset}$ is subject to impure squeezing. 
The discrepancy with eq.~\eqref{eq:LossPreservation} is considered to be coming from the annihilation operator replaced by the single-mode one $\hat{a}(t)\to\hat{a}$. 
In fact, 
\begin{align}
\hat{a}_g\hat{P}_r^\dagger\ket{\emptyset} 
\propto & \Bigl[\langle g,N(g^\ast \ast r)\rangle \hat{a}_g^\dagger \notag\\
& + \sqrt{1-\lvert\langle g,N(g^\ast \ast r)\rangle\rvert^2}\hat{a}_{g^\perp}^\dagger\Bigr]\ket{\emptyset}, 
\end{align}
and this coincides with the wrong answer of the lossy single-photon state in the wavepacket mode $g(t)$. 
We must be careful about this way of wrong consideration, coming from the replacement of an instantaneous annihilation operator $\hat{a}(t)$ by a single-mode one $\hat{a}$, which may especially occur when orthogonal modes are traced out at the beginning.

As for the impure squeezing of the heralded single photon state, the calculated $10\%$ as the mode-mismatch is rigorous when $r(t) \propto \exp(-\gamma \lvert t\rvert)$, but we note that the actual portion of ill photon pairs is not exactly $10\%$ for the following reason. 
The precise calculation must include the effect to $\sqrt{n+1}$ terms which appears when $\hat{a}^\dagger$ is applied to the $n$-photon state $\ket{n}$ (i.e., the effects that photons tend to bunch due to stimulated processes), which makes the situations much more complicated. 
In the case of weak pumping where $\hat{S}_r\approx1+(1/2)(\hat{P}_r^\dagger+\hat{P}_r)$ and the higher-order terms like $\hat{P}_r^{\dagger2}$ are negligible, 
\begin{align}
& \bigl[c_{g,g}\hat{a}_g^{\dagger2} + 2c_{g,g^\perp}\hat{a}_g^\dagger\hat{a}_{g^\perp}^\dagger\bigr]\ket{1}_g\otimes\ket{0}_{g^\perp} \notag\\
= & \sqrt{6}c_{g,g}\ket{3}_g\otimes\ket{0}_{g^\perp} + 2\sqrt{2}c_{g,g^\perp}\ket{2}_g\otimes\ket{1}_{g^\perp}. 
\end{align}
That is, the process of $\ket{1,0}\to\ket{3,0}$ is three times more significant than the process of $\ket{1,0}\to\ket{2,1}$, but taking into account the factor $2$ multiplied to $c_{g,g^\perp}$ in the decomposition of $\hat{P}_r^\dagger$ in eq.~\eqref{eq:SpecificWavepacketPair} or eq.~\eqref{eq:WavepacketSeries}, the actual bias in the ratio is $(\sqrt{6})^2:(2\sqrt{2})^2=3:4$. 
Anyway, the above discussion has shown the existence of the inherent inefficiency in the conventional photon-subtraction method, due to the mode-mismatch of squeezing with respect to the heralded single-photon mode. 
Next, we discuss how this inefficiency can be removed.

\section{Purification by filtering}
\label{sec:Filter}

As discussed above, under the time-translation symmetry of photon-pair generation with no optical losses, each frequency component of squeezing is pure. 
That is, by limiting the bandwidth of wavepackets, the squeezing of the wavepackets becomes purer. 
Therefore, if we could create cat states in wavepackets with a narrower bandwidth (relative to the bandwidth of squeezing), the mode-match of squeezing is improved and thereby the resulting cat states become purer.

Here we propose the method to insert a filter cavity before the photon detection, which has narrower bandwidth than that of the OPO. 
In the following, we will show that the mode-matching rate of squeezing can arbitrarily approach to $1$ with this method. 
Therefore, our method enables the ideal squeezed single-photon state, which was not possible with the conventional method. 
It is in contrast to previous experiments where, although filter cavities are utilized, the bandwidths of them are broader than that of the OPO in order to utilize the natural photon-pair correlations determined by the OPO cavity as described above.

Expressing the response function of the filter as $h(t)$, the transformation of the field passing through the filter is defined as, 
\begin{align}
& \hat{F}_h^\dagger\hat{a}(t)\hat{F}_h \notag\\
= & \int \Bigl\{h^\ast(\tau)\hat{a}(t-\tau) \notag\\
& - [\delta(\tau)-h^\ast(\tau)]\hat{a}_\text{ref}(t-\tau)\Bigr\}d\tau, 
\end{align}
where the subscript `ref' denotes the reflected ancillary field to compensate the commutation relation. 
The exact form of the response function of a typical single-cavity filter is the same as the impulse response function of a first-order low-pass filter, 
\begin{align}
h(t) = \Gamma\exp(-\Gamma t)u(t). 
\end{align}
This is equivalent to a frequency-dependent beamsplitter \cite{Collet.PRA1984}, 
\begin{align}
\hat{F}_h^\dagger\hat{\widetilde{a}}(\omega)\hat{F}_h 
= & \frac{\Gamma\hat{\widetilde{a}}(\omega)+i\omega\hat{\widetilde{a}}_\text{ref}(\omega)}{\Gamma-i\omega}, 
\end{align}
where the best transmission is obtained at the center frequency of the cavity Lorentzian $\omega=0$. 
Here, $\Gamma$ is the filter-cavity decay rate (with the factor 2). 
It can be checked $h(t)\to\delta(t)$ at the limit of $\Gamma\to0$, corresponding to the case without any filtering, $\hat{F}_h^\dagger\hat{a}(t)\hat{F}_h\to\hat{a}(t)$. 
However, for the moment, we deal with the filter response function $h(t)$ as a general function, allowing the filter to be a more general one.

Modifying eqs.~\eqref{eq:Subtraction} and \eqref{eq:SubtractionLimit} to the beam version, and using the invariance of a vacuum state under filtering, we obtain, 
\begin{align}
& \,_\text{ref}\bra{\emptyset}\otimes \,_\text{anc}\bra{\emptyset}\hat{a}_\text{anc}(t)\hat{F}_{h;\text{anc}}\hat{B}(R\to0) (\hat{S}_r\ket{\emptyset})\otimes\ket{\emptyset}_\text{anc}\otimes\ket{\emptyset}_\text{ref} \notag\\
= & \,_\text{ref}\bra{\emptyset}\otimes \,_\text{anc}\bra{\emptyset} \notag\\
& \Bigl[ \int \Bigl\{h^\ast(\tau)\hat{a}_\text{anc}(t-\tau) - [\delta(\tau)-h^\ast(\tau)]\hat{a}_\text{ref}(t-\tau)\Bigr\}d\tau\Bigr] \notag\\
& \hat{B}(R\to0) (\hat{S}_r\ket{\emptyset})\otimes\ket{\emptyset}_\text{anc}\otimes\ket{\emptyset}_\text{ref} \notag\\
= & \sqrt{R} \Bigl[\int h^\ast(\tau)\hat{a}(t-\tau)d\tau\Bigr] \hat{S}_r\ket{\emptyset}. 
\end{align}
Here, eq.~\eqref{eq:SubtractionLimit} is applied to $_\text{anc}\bra{\emptyset}\hat{a}_\text{anc}(t-\tau)\hat{B}(R\to0)(\hat{S}_r\ket{\emptyset})\otimes\ket{\emptyset}_\text{anc}$. 
Note that, in addition to the necessary projection measurement $_\text{anc}\bra{\emptyset}\hat{a}_\text{anc}(t)$, here $_\text{ref}\bra{\emptyset}$ is also applied for mathematical simplicity, but actually the measurement of the reflected field is not needed.

We set the photon detection timing to be $t = 0$ without loss of generality. 
Then, the photon subtraction after filtering by $h(t)$ is to apply the annihilation operator $\hat{a}_{N(h^R)}$. 
The squeezed single-photon state in eq.~\eqref{eq:SqueezedSinglePhotonCW} is modified as, 
\begin{align}
& \hat{a}_{N(h^R)}\hat{S}_r\ket{\emptyset} \notag\\
& = \Bigl[\int N(\widetilde{h}^{R\ast})(\omega)\hat{\widetilde{a}}(\omega)d\omega\Bigr]\hat{S}_r\ket{\emptyset} \notag\\
& = \hat{S}_r\Bigl[\int N(\widetilde{h}^{R\ast})(\omega)\hat{\widetilde{a}}^\dagger(\omega)\exp\{2i\theta(\omega)\}\sinh \lvert\widetilde{r}(\omega)\rvert d\omega\Bigr]\ket{\emptyset} \notag\\
& \approx \hat{S}_r\Bigl[\int N(\widetilde{h}^{R\ast})(\omega)\widetilde{r}(\omega)\hat{\widetilde{a}}^\dagger(\omega)d\omega\Bigr]\ket{\emptyset} \notag\\
& \propto \hat{S}_r\hat{a}_{N(h^{R\ast} \ast r)}^\dagger\ket{\emptyset}. 
\end{align}
The wavepacket of the heralded single-photon component is modified from $N(r)(t)$ in eq.~\eqref{eq:GeneralWavepacket} to $N(h^{R\ast}\ast r)(t)$ by the filtering with $h(t)$. 
Therefore, the concerned mode-matching rate is, 
\begin{align}
M[h^R\ast r, r] = & \lvert \langle N(h^R\ast r),N(h^R\ast r\ast r)\rangle \rvert^2. 
\end{align}
In the extreme case where $r(t)$ almost works like a delta function in convolution with $h^R(t)$ except for the normalization, $M[h^R\ast r, r]$ approaches to unity because $N(h^R)(t) \approx N(h^R \ast r)(t) \approx N(h^R \ast r \ast r)(t)$. 
This is the situation we aim at by inserting the filter.

\begin{figure}[tb]
\centering
\includegraphics{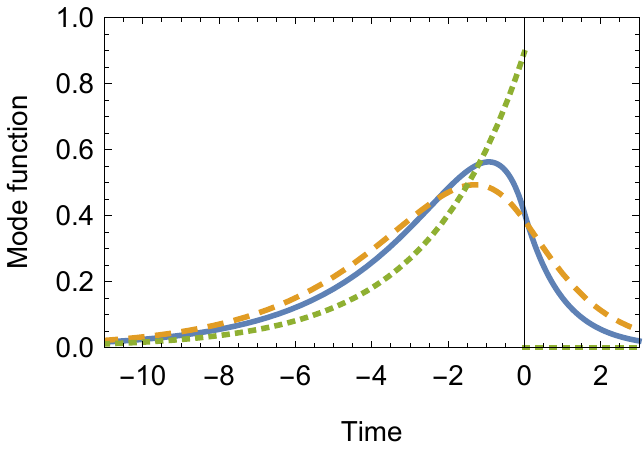}
\caption{
(Color online) 
Mode functions to show the mode-match of squeezing to heralded single-photon wavepackets, for the case with filtering, for $r(t)\propto\exp(-\gamma\lvert t\rvert)$ with $\gamma=1$ and $h(t)\propto\exp(-\Gamma t)u(t)$ with $\Gamma=0.4$. 
Solid blue: $N(h^R \ast r)(t)$. 
Dashed orange: $N(h^R \ast r\ast r)(t)$. 
Dotted green: $N(h^R)(t)$. 
}
\label{fig:ModeFuncFiltered}
\end{figure}

Now we consider the specific case of $h(t) = h^\ast(t) \propto \exp(-\Gamma t)u(t)$ and $r(t) = r^\ast(t) \propto \exp(-\gamma\lvert t\rvert)$, and see the mode-matching rate improved by the filtering. 
The heralded single-photon wavepacket mode $N(h^R\ast r)(t)$ is calculated as, 
\begin{align}
& N(h^R\ast r)(t) \notag\\  
= & 
\begin{cases}
\sqrt{\frac{\gamma\Gamma}{2\gamma+\Gamma}}\frac{1}{\gamma-\Gamma}\Bigl[2\gamma\exp(\Gamma t) - (\gamma+\Gamma)\exp(\gamma t)\Bigr], & t<0, \\
\sqrt{\frac{\gamma\Gamma}{2\gamma+\Gamma}}\exp(-\gamma t), & t\ge0, 
\end{cases}
\end{align}
if $\gamma\neq\Gamma$. 
On the other hand, $N(h^R\ast r\ast r)$ is calculated as, 
\begin{align}
& \sqrt{\frac{16\gamma^3+29\gamma^2\Gamma+20\gamma\Gamma^2+5\Gamma^3}{2\gamma^3\Gamma}}N(h^R\ast r\ast r)(t) \notag\\
= & 
\begin{cases}
\frac{4\gamma^2}{(\gamma-\Gamma)^2}\exp(\Gamma t) - \Bigl[\frac{(2\gamma-\Gamma)(\gamma+\Gamma)^2}{\gamma(\gamma-\Gamma)^2}-\frac{(\gamma+\Gamma)^2}{\gamma-\Gamma}t\Bigr]\exp(\gamma t) & t<0, \\
\Bigl[2+\frac{\Gamma}{\gamma}+(\gamma+\Gamma)t\Bigr]\exp(-\gamma t) & t\ge0, 
\end{cases}
\end{align}
if $\gamma\neq\Gamma$. 
Figure~\ref{fig:ModeFuncFiltered} shows the three functions $N(h^R\ast r)(t)$, $N(h^R\ast r\ast r)$, and $N(h^R)$, for $\gamma = 1$ and $\Gamma = 0.4$. 
In comparison with the functions without filtering shown in Fig.~\ref{fig:ModeFuncBare}, we can see that the heralded wavepacket mode $N(h^R\ast r)(t)$ approaches to a rising exponential wavepacket mode $N(h^R)$, and the mode overlap with $N(h^R\ast r\ast r)$ is improved.

\begin{figure}[tb]
\centering
\includegraphics{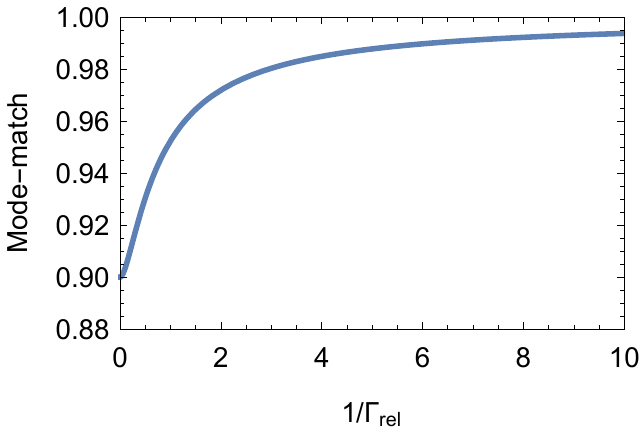}
\caption{The mode-matching rate in eq.~\eqref{eq:ModeMatchFiltered}.}
\label{fig:ModeMatch}
\end{figure}

From above, the mode-matching rate is calculated as a function of the relative bandwidth of the filter $\Gamma_\text{rel} \coloneqq \Gamma/\gamma$, 
\begin{align}
M[h^R\ast r, r] = \frac{(8+9\Gamma_\text{rel}+3\Gamma_\text{rel}^2)^2}{2(2+\Gamma_\text{rel})(16+29\Gamma_\text{rel}+20\Gamma_\text{rel}^2+5\Gamma_\text{rel}^3)}. 
\label{eq:ModeMatchFiltered}
\end{align}
It approaches to the bare mode-matching rate of $9/10$ in eq.~\eqref{eq:BareModeMatch} at the limit of $\Gamma_\text{rel}\to\infty$ and to unity at the limit of $\Gamma_\text{rel}\to0$. 
The mode-matching rate $M[h^R\ast r, r]$ is plotted in Fig.~\ref{fig:ModeMatch} with respect to the relative inverse bandwidth $1/\Gamma_\text{rel}$. 
We can see that $M[h^R\ast r, r]$ monotonically improves with larger $1/\Gamma_\text{rel}$. 
That is, narrower bandwidth of the filter compared with that of the OPO is preferable regarding pure cat-state creation. 
In the same way, we can also consider a higher-order low-pass filter by combining multiple cavities, in which case the mode-matching rate will approach to unity more rapidly.

\section{Summary and Discussion}
\label{sec:summary}

We discussed inherent impurity in the conventional CW-based photon subtraction methods due to nonflat spectrum of OPO squeezing. 
The impurity was characterized via the notion of mode-mismatch of squeezing, with highlighted eq.~\eqref{eq:SpecificWavepacketPair}. 
Then we showed that the impurity is arbitrarily reduced by inserting a filter cavity before the photon detection for the photon subtraction. 
The amount of inefficiencies discussed here and removed by our filtering method may not be so large, but it will become important for ultimate experiments where very high purities of cat states are required.

We here basically discussed one-photon subtraction, but the same discussions are valid for multiphoton subtraction. 
However, in the case of multiphoton subtraction, there arise additional parameters of time differences among individual photon detections \cite{Takahashi.PRL2008,Takeoka.PRA2008}.

Since the impurity is coming from the longitudinally continuous nature of the squeezed light, making entanglement with orthogonal modes, another possible solution to remove the impurity is first to create a pure single-mode squeezed state directly inside an ideal quantum memory and then to subtract photons from the pure squeezed state released from the memory. 
However, currently the inefficiency of a cavity-based quantum memory itself is much larger than the inefficiency discussed here \cite{Yoshikawa.PRX2013}.

Additional observation is, althought here we only discussed the CW pumping case, we expect there will be similar problems when the parametric down conversion is implemented with pulsed pump laser light. 
In the pulsed case, highest level of squeezing will be available around the peaks of pump pulses, while relatively low squeezing will exist at the side slopes \cite{Eto.OptExpr2008}. 
This situation would naturally involve multimode squeezing, which leads to impure photon subtraction.

In general, optical filtration before heralding-photon detection is useful, enabling engineering of the heralded wavepacket modes, and we note that optical high-pass filtering is utilized in previous experiments for preparation of cat states in order to avoid noisy low-frequencies, which are then utilized as input states of quantum teleportation system \cite{Takeda.PRA2012}.

\section*{Acknowledgment}

This work was partly supported by CREST of JST, JSPS KAKENHI, and APSA, of Japan.

\end{document}